\documentclass[12pt]{article}

\usepackage{fullpage} 
\usepackage[margin=2.5cm,top=2cm,bottom=3cm,centering]{geometry}
\usepackage{graphicx,graphics}
\usepackage[version=3]{mhchem}
\usepackage{amsmath} 
\usepackage{amsfonts}
\usepackage{multicol}

\title{New treatments of density fluctuations and recurrence times for re-estimating Zermelo's paradox}

\author{Denis Michel\\
\\
       Universite de Rennes1-IRSET. Campus de Beaulieu Bat. 13.\\
        35042 Rennes cedex. \ denis.michel@live.fr}

\date{} 

\begin{document}
\maketitle
\textbf{Abstract}
What is the probability that all the gas in a box accumulates in the same half of this box? Though amusing, this question underlies the fundamental problem of density fluctuations at equilibrium, which has profound implementations in many physical fields. The currently accepted solutions are derived from the studies of Brownian motion by Smoluchowski, but they are not appropriate for the directly colliding particles of gases. Two alternative theories are proposed here using self-regulatory Bernoulli distributions. A discretization of space is first introduced to develop a mechanism of matter congestion holding for high densities. In a second mechanism valid in ordinary conditions, the influence of local pressure on the location of every particle is examined using classical laws of ideal gases. This approach reveals that a negative feedback results from the reciprocal influences between individual particles and the population of particles, which strongly reduces the probability of atypical microstates. Finally, a thermodynamic quantum of time is defined to compare the recurrence times of improbable macrostates predicted through these different approaches.
\newline
\newline
\noindent

\newpage
\section{Introduction}
If you stand in one side of a closed room and that all the air suddenly accumulates on the other side, you would fatally suffer from depressurization. Yet nobody seems afraid by this horrible scenario. People without notions of statistical mechanics just feel that this is impossible. Statistical physicists consider that this situation is improbable, but so improbable that it is completely negligible. In fact, this probability has never been exactly calculated. This question touches the long standing paradox of macroscopic transitions considered at first glance as irreversible but underlain by microscopic reversibility. New calculations are presented in this article to quantify the recurrences times of Poincar\'e.

\section{Macro-irreversibility resulting from micro-reversible processes}
According to the second law of thermodynamics, macrosystem evolutions can be classified into reversible and irreversible ones. From a statistical perspective, irreversible transitions transform macrostates covering a lower number of microstates (less probable) into macrostates covering a higher number of microstates (more probable). A simple example is the expansion in a box of a gas previously confined in a small subvolume after removal of a partition membrane. This rational principle of irreversibility however raised concerns, particularly vehement from Zermelo \cite{Zermelo}. For historical reviews on these debates, see \cite{Steckline,Brown} and the interesting articles of Boltzmann translated and compiled in \cite{Brush}. The Loschmidt reversibility paradox \cite{Loschmidt} stressed that when microscopic changes are reversible, there is no reason for an ensemble of microscopic changes to be not reversible. This view is supported by the recurrence theorem of Poincar\'e, which states that any very improbable state, such as a well organized initial state, will necessarily reappear, given sufficient time \cite{Poincare}. Considering the example of gas expansion described above, such an event would be the spontaneous reclustering of gas particles in a small region of the box. Moreover, as this state will reappear once, it will reappear infinitely often, giving rise to a quasi-periodic phenomenon. The right question is not whether this event can occur or not, but how often it reproduces. As pointed by Zermelo, this paradox ruins the idea of irreversibility. The answer of Boltzmann to Zermelo was: "\textit{Young man, you know mathematics but you do not know physics. Think how long it takes for that to happen? It would take far longer than the age of the universe even for a very small system}". For his calculus, Boltzmann used the example of 1 cm$ ^{3} $ of gas with the assumption that the system passes through all its microstates before recurrence and that $ 2\times 10^{27} $ different microstates take place in one second \cite{Boltzmann}. However, Boltzmann enumerated only microstates differing through their velocity, but not spatial, distribution, although density changes are much more appreciable macroscopically. This puzzling choice could be merely technical because of the difficulty to enumerate spatial configurations in continuous space. On the contrary, changes of particle location, but not of velocities, will be examined here, by assuming that velocity vectors distribute evenly on particles regardless of their instantaneous coordinates. The time unit of universe age used by Boltzmann will be reused. Indeed, the disconnection between realistic observation time windows and the recurrence waiting times remains an excellent argument opposable to Zermelo, but in fact the parameter of time is absent from microstate enumerations and this point has never been really quantified. In addition, the theory generally believed to address density fluctuations, that of Smoluchowski, is inappropriate to quantify recurrence periods for self-colliding systems such as gases. 

\section{Traditional modeling}
A traditional tool of statistical mechanics is a box of volume $ V $ containing $ N $ randomly distributed particles. In this box, the number of particles $ x $ present in a subvolume $ v $, is subject to fluctuations around a mean value $ Nv/V $. The probability that $ x=n $ is currently given by a simple Bernoulli distribution:

\begin{equation} P(x=n) = \binom{N}{n} \left (\frac{v}{V}\right )^{n} \left (1-\frac{v}{V}\right )^{N-n} \end{equation} 
\noindent
which says that the probability of location of a single particle is dictated by the fraction of available volumes only \cite{Landau}. This admitted result can however be questioned because it neglects both particle exclusion and pressure and is valid only for very low densities. The available space is more and more reduced as density increases. As a consequence, Eq.(1) overestimates the probability of atypical microstates such as those in which near all the particles are in the same side of the box. Two mechanisms are proposed below to restore a role for the relative particle density in the compartments: (\textbf{i}) the bulk or congestion hypothesis, only appreciable for highly concentrated systems, and (\textbf{ii}) the pressure hypothesis, important for gases in ordinary conditions and missing in the current approaches to density fluctuations. But let us first point out the assumptions of Smoluchowski which are inappropriate for gases.

\section{Brownian motion vs mutual collision}
The study of Smoluchowski on density fluctuations in a medium of constant average density \cite{Smoluchowski}, is far from general. Smoluchowski's calculations were celebrated because they remarkably fitted the experiment of Svedberg on particles moving through Brownian motion \cite{Svedberg}. But Brownian motion does not result from mutual collisions. The difference is important since the main starting assumption of Smoluchowski is that the motions of the different particles are not mutually influenced, so that all possible individual positions have equal \textit{a priori} probabilities. With respect to the problem of density fluctuations, this assumption implies that the probability of entrance of particles in any subvolume, does not depend on the number of particles already contained in it. Accordingly, the basic ingredient used by Smoluchowski was a Poisson distribution $ P(x=n)= \nu ^{n} \textup{e} ^{-\nu}/n! $ corresponding to the limit of the binomial formula of Eq.(1) for very large $ N $ and $ V $ and a mean value $\nu = Nv/V $. The movements of colloids are the resultant of huge numbers of collisions with solvent molecules but not of direct collisions between the suspended substances. Moreover, the motion of these substances is buffered by surrounding solvent molecules, so that they are not transferred to other suspended particles, thereby ensuring the mutual independence of these particles and their insensitivity to crowding and pressure. This independence no longer applies to systems made of mutually colliding particles. Although gas particles are usually called independent when non-interacting chemically, they are in fact mutually constrained. A very interesting consequence of this situation is that every individual particle is both (\textbf{i}) an actor of the collective effect known as pressure and (\textbf{ii}) a receptor of this pressure which influences its own location. These reciprocal influences between the individual and the population is the essence of statistical mechanics and should appear in the formulation of density fluctuations, contrary to Eq.(1) and to the assumptions of Smoluchowski.

\section{The congestion mechanism for dense systems}
The description of this mechanism necessitates discretizing space. The discretization of space allows to reconcile the classical approaches to statistical dynamics with the principle of spatial exclusion. Indeed, the pioneer versions of the ideal gas theories ignored the size of the particles, which were conceived as superposable points whose number is defined in $ \mathbb{N} $, scattered in continuous volumes defined in $ \mathbb{R}^{3} $. Hence, considering $ V^{N}/N! $ as a number of microstates, as often found in the literature, is formally erroneous. This problem can be solved by discretizing space.

\subsection{Quantum of space}
The concepts of volume per particle and of mutual exclusion have been introduced in more elaborated versions of kinetic theories, for instance to describe transportation properties such as viscosity, diffusion and thermal conductibility. Moreover, a thermal Planck volume has been defined as the volume necessary to accommodate a single particle at given temperature, shape and mass, with a mean momentum $ p $. For ideal gas particles, $ \lambda $ is the thermal wavelength of de Broglie $ \lambda = h/p =  h/\sqrt{2\pi m k_{B}T} $ where $ h $ is the Planck constant, $ k_{B} $ the Boltzmann constant and $ T $ the temperature. This elementary volume allows to define discrete volumes corresponding to the integers closer to $ \acute{V}=V/\lambda ^{3} $. The $ \lambda ^{3} $ parcel is not an intrinsic property of "empty space" but is defined relatively to the occupying particles, which explains why it depends on temperature. The same space discretization is in fact implicit in the Sackur-Tetrode equation for entropy \cite{Pathria}, which introduced quantum considerations in classical thermodynamics and has the advantage to improve entropy extensivity in the Gibbs experiment. As shown below, this discretization is also useful for calculating fluctuation densities of concentrated systems for which the volume occupied by the particles cannot be neglected.

\subsection{Fluctuation probabilities of dense systems}
This theory applies to any dense system, including suspensions. If the volumes $ v $ and $ V $ contain $ n $ and $ (N-n) $ particles respectively, the probability for a $ (N+1) $th particle to be in $ v $ is not $ v/V $ as assumed in Eq.(1), but $ (v-n\lambda ^{3})/(V-N\lambda ^{3}) $. This difference is of course negligible when $ V $ is very large compared to $ N\lambda ^{3} $, but not for dense systems. When the intrinsic volume of particles is no longer negligible, the general probability that $ n $ particles are present in $ v $, can be obtained using the unitless volumes $ \acute{v} $ defined above. This discretization does not forbid using Eq.(1), rewritten for the circumstance
\begin{equation}  P(x=n) = \binom{N}{n} \frac{\acute{v}^{n}(\acute{V}-\acute{v})^{N-n}}{\acute{V}^{N}} \end{equation}

In this equation, the exponents mean that the entry of a particle in one compartment does not depend on the number of particles already in this compartment. But in fact, when one particle is present in $ v $, then $ \acute{v}-1 $ places are free; when 2 particles are present, $ \acute{v}-2 $ places remain available etc. Hence, Eq.(2) becomes

\begin{equation} P(x=n) = \binom{N}{n}\frac{[\acute{v}(\acute{v}-1)...(\acute{v}-n+1)][(\acute{V}-\acute{v}) (\acute{V}-\acute{v}-1)...(\acute{V}-\acute{v}-N+n+1)]}{\acute{V}(\acute{V}-1)...(\acute{V}-N+1)} \end{equation}
\noindent 
that can be rewritten as
\begin{equation} P(x=n) = \binom{\acute{v}}{n}\binom{\acute{V}-\acute{v}}{N-n}/\binom{\acute{V}}{N} \end{equation}

Note that instead of starting from Eq.(1), the result can be conceived directly using the hypergeometric distribution of Eq.(4) which describes the number of ways to distribute $ n $ and $ (N-n) $ particles in $ \acute{v} $ and $ (\acute{V}-\acute{v})$ cells respectively, among a total number of possible distributions of $ N $ particles over $ \acute{V} $ cells. Eq.(4) is intrinsically normalized, as should be a probability density function, owing to the Chu-Vandermonde identity:

\begin{equation} \sum_{i=0}^{k}\binom{a}{i}\binom{b}{k-i}=\binom{a+b}{k} \end{equation}

One can verify that Eq.(1) is recovered when the particles are very scarce in a very large volume. Indeed, Eq.(3) can be rearranged as

\begin{equation} P(x=n) = \binom{N}{n}\frac{\acute{v}^{n}(\acute{V}-\acute{v})^{N-n}}{\acute{V}^{N}} R \end{equation} 
with 
\begin{equation} R=\frac{\left [\left (1-\frac{1}{\acute{v}}  \right )\left (1-\frac{2}{\acute{v}}  \right ) ...\left (1-\frac{\acute{v}-n+1}{\acute{v}}  \right ) \right ]\left [\left (1-\frac{1}{\acute{V}-\acute{v}}  \right )\left (1-\frac{2}{\acute{V}-\acute{v}}  \right ) ...\left (1-\frac{N-n+1}{\acute{V}-\acute{v}}  \right ) \right ]}{\left (1-\frac{1}{\acute{V}}  \right )\left (1-\frac{2}{\acute{V}}  \right ) ...\left (1-\frac{N+1}{\acute{V}}  \right )}  \end{equation} 

If $ \acute{V} >> N $, it is likely that $ \acute{v} >> n $ and then that $ (\acute{V}-\acute{v}) >> (N-n) $, so that $ R \sim 1 $. Then, Eq.(6) reduces to Eq.(2) and to Eq.(1) since the $ \lambda ^{3} $ factor disappears.

A satisfactory behavior of Eq.(4) is that the probabilities of typical configurations is higher than for Eq.(1), while those of highly asymmetric  configurations are much lower (Fig.1). However, this difference is only marginal for moderately dense systems since the variance of the distribution of Eq.(4) is only $ 1-\dfrac{N\lambda ^{3}}{V} $ times that of Eq.(1). A much larger difference can be obtained by introducing a role for local pressure in particle distribution.
\newline
\begin{center}
\includegraphics[width=12cm]{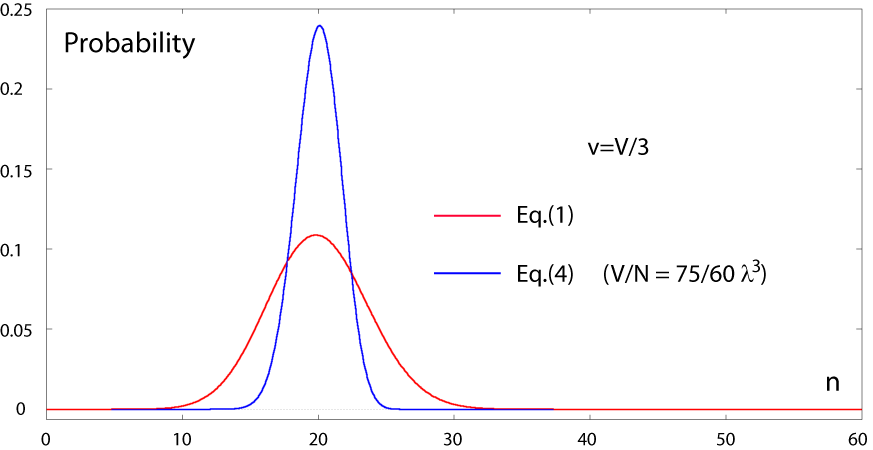}
\end{center}

\begin{small}
\textbf{Figure 1}. Comparative results of Eqs(1) and (4). For a small but dense system made of 60 particles running and colliding in a box of $ V/\lambda ^{3} $ = 75 cells, the curves represent the probabilities of finding $ n $ particles in a subvolume of 25 cells, calculated using either Eq.(1) (red) or Eq.(4) (blue, darker). Eq.(1) underestimates the symmetric microstates around the mean density $ n $ = 20 ($ n=Nv/V) $ and conversely overestimates the probability of the asymmetric microstates, because it does not take into account the intrinsic volumes of the particles.
\end{small}

\section{The pressure feedback}
Suppose that a box contains $ N $ particles, out of which $ n $ are present in a subvolume  $ v $ and $ (N-n) $ are in the remaining volume $ (V-v) $. As previously, these compartments are defined conceptually and are not enclosed by any physical boundary. Now imagine the fate of a single particle present at the interface between $ v $ and $ (V-v) $. To determine the probability that this particle enters either $ v $ or $ (V-v) $, it is reasonably not acceptable to neglect the influence of the relative densities in these subvolumes, because of the associated probability of expulsive collisions. This parameter is related to pressure. A crude modeling is proposed to illustrate the consequence of incorporating this parameter, as follows. Pressure represents the density of collisions, itself proportional to the density of the particles ($ \frac{N}{V} $) at given temperature (because $ P= \frac{N}{V}k_{B}T $). The probability (written below $ P[\in ...] $) for any particle to enter one compartment rather than the other one, is a function of the competitive propensities of these compartments to recruit this particle. These propensities are named $ F_{v} $ and $ F_{V-v} $ for the compartments $ v $ and $ (V-v) $ respectively. Given that they should be normalized to 1 over the whole box, the probabilities take the forms

\begin{subequations} \label{E:gp}
\begin{equation} P[\in v] =\frac{F_{v}}{F_{v}+F_{V-v}} \end{equation} \label{E:gp1}
and 
\begin{equation} P[\in (V-v)] =\frac{F_{V-v}}{F_{v}+F_{V-v}} \end{equation} \label{E:gp2}
\end{subequations}
\noindent
The propensity to incorporate a particle should now be defined. It seems rational to postulate that $ F $ is the product of two parameters which concur equally to take up a particle, which are:

\begin{itemize}
\item The fractional volume available in the absorbing compartment (the only parameter considered in Eq.(1)), whose values are $ \frac{v}{V}  $ for the compartment $ v $ and $ \left (1-\frac{v}{V}  \right ) $ for the compartment ($ V-v $).
\item The fractional pressure in the rejecting compartment, equivalent to the outside fractional particle density, whose values are here $ \frac{(N-n)/(V-v)}{N/V}= \left (1-\frac{n}{N}  \right )/\left (1-\frac{v}{V}  \right )  $ for the compartment $ v $ and $ \frac{n/v}{N/V}= \left (\frac{n}{N}  \right )/\left (\frac{v}{V}  \right ) $ for the compartment $ (V-v) $.
\end{itemize}

Introducing these values into Eq.(8) gives

\begin{subequations} \label{E:gp}
\begin{equation} P[\in v] = \frac{\left(1-\frac{n}{N} \right )\left(\frac{v}{V}\right)^{2}}{\left(1-\frac{n}{N} \right )\left(\frac{v}{V}\right)^{2}+\left(\frac{n}{N} \right )\left(1-\frac{v}{V}\right)^{2}} \end{equation} \label{E:gp1}
and 
\begin{equation} P[\in (V-v)] = \frac{\left(\frac{n}{N} \right )\left(1-\frac{v}{V}\right)^{2}}{\left(1-\frac{n}{N} \right )\left(\frac{v}{V}\right)^{2}+\left(\frac{n}{N} \right )\left(1-\frac{v}{V}\right)^{2}} \end{equation} \label{E:gp2}
\end{subequations}

These equations satisfactorily simplify in singular situations. If the density in $ v $ is the same as in $ (V-v) $, that is to say when $ n=Nv/V $, one obtains

\begin{subequations} \label{E:gp}
\begin{equation} P[\in v] =\frac{v}{V} \end{equation} \label{E:gp1}
and 
\begin{equation} P[\in (V-v)] =1-\frac{v}{V} \end{equation} \label{E:gp2}
\end{subequations}

\noindent
which are simply the probabilities used in Eq.(1). Reciprocally, if the compartments have the same volume ($ v=V/2 $), then the probabilities become only dependent on the number of particles in these compartments, as expected.

\begin{subequations} \label{E:gp}
\begin{equation} P[\in v] =1-\frac{n}{N} \end{equation} \label{E:gp1}
and 
\begin{equation} P[\in (V-v)] =\frac{n}{N} \end{equation} \label{E:gp2}
\end{subequations}

\noindent
This result, symmetric to the previous one found when the number of particles does not matter, is acceptable for fluctuation ranges of ordinary macroscopic systems. The next step of the reasoning is to understand that every individual particle can be considered as the $ N $th component of a population already containing all others, so that all the particles are equally submitted to Eq.(9) and are functions of $ n $ ranging from 0 to $ N $ and $ v $ ranging from 0 to $ V $. Finally, a factor 2 is introduced for normalization of the probability that $ n $ particles are present in fixed $ v $, yielding the completely developed equation

\begin{equation} P(x=n) = 2\binom{N}{n}\left (\frac{(N-n)v^{2}}{(N-n)v^{2}+n(V-v)^{2}}  \right )^{n} \left (\frac{n(V-v)^{2}}{(N-n)v^{2}+n(V-v)^{2}} \right )^{N-n} \end{equation}

\noindent
\newline
This equation is a self-regulatory Bernoulli distribution embodying the circular influence between individuals and the population, which is a negative feedback. This negative feedback severely readjusts any deviation from equipartition back to equipartition, with a strength even stronger when the gap is large. Numerical application reveals that Eq.(12) tolerates very much less than Eq.(1) atypical microstates such as strongly asymmetric ones containing few particles in $ v $ or $ (V-v) $. To concretely conceive this point, if all the particles are already present in the left half of the box (a purely theoretical hypothesis), then the probabilities for an additional particle to enter either the left or right halves are not 1/2 and 1/2 as in Smoluchowski's theory, since it will surely bounce back in the empty side.

\section{Recurrence times}
Strongly asymmetric distributions are supposed to be so improbable that they never occur spontaneously. Physically speaking, the recurrence time of Poincar\'e of extreme microstates is expected to exceed all possible observation windows. However, to formally establish this point, it is first necessary to define the time step separating two different configurations in a box of gas. Indeed, if particle trajectories are conceived as continuous, then any time window $ \delta t $, as short as desired, is expected to contain an infinite number of configurations. As noticed by Zermelo, probabilities do not contain temporal terms. This problem forbids the calculation of recurrence times and contradicts the finite number of configurations assumed in the Sackur-Tetrode equation. Once again, discretization (in this case of time) can save us.

\subsection{Thermal quantum of time}
The goal is now to define the time step $ \tau $ separating two consecutive configurations. Transitions between successive states of the system were previously assumed to correspond to collisions \cite{Rice}. If collisions can indeed discretely punctuate velocity changes, they are not necessary for location changes. Instead, the quantum of time proposed here is the time necessary to cross the length unit below which successive configurations cannot be distinguished, because of the uncertainty principle. This is precisely what is the thermal wavelength of de Broglie $ \lambda $. Hence, for a mean particle velocity $ \left \langle v \right \rangle = \sqrt{8k_{B}T/\pi m} $ obtained by averaging the velocities distribution \cite{Maxwell}, one obtains

\begin{equation} \tau = \frac{\lambda }{\left \langle v \right \rangle}= \frac{h}{4 k_{B}T} = 4\times 10^{-14} \ \textup{s} \ \textup{at} \ T=300 \ \textup{K} \end{equation}
\noindent
Recurrence times can now be calculated.

\subsection{Calculation of recurrence times}

\subsubsection{For a microstate}
A microstate of probability $ P = 1/\Omega $, is expected to appear after a mean waiting time $ \left \langle T \right \rangle = \Omega  \tau $, regardless of the initial observation time.

\subsubsection{For a macrostate}
A macrostate covers a certain number of microstates giving the same macroscopic outcome $ X $; for example those in which all the particles are in the same half of a box. The probability of a macrostate $ X $ is the fraction of these microstates $ P_{X} = \Omega _{X}/\Omega $. The mean recurrence time of a macrostate is of course lower than those of its constituent microstates, such that

\begin{equation}  \left \langle T \right \rangle _{X} = \dfrac{\Omega}{\Omega _{X}}  \tau  = \dfrac{\tau}{P_{X}} \end{equation}

Strictly speaking, $ \left \langle T \right \rangle _{X} $ is not a period because rare events are not regularly spaced in time. Given the memoryless nature of transitions resulting from numerous chaotic collisions \cite{Michel}, the probability that a given state occurs before $ t $ is

\begin{equation}  P(X \leq  t) = 1- \textup{e}^{t/\left \langle T \right \rangle_{X}} \end{equation}

To get a concrete idea of the potency of Eq.(12) compared to Eq.(1), the recurrence times expected from the classical and the pressure approaches, are compared in the case of the distribution of only 100 gas particles (Table 1). Results are obtained by integrating the probability density functions of Eqs(1) and (12).

\begin{table}[!h]
\caption{Recurrence times of atypical macrostates in which more than 90 out of $ 100 $ particles are confined in 3/4 or 1/2 box. The results shown are obtained with the classical approach (Eq.(1)) and with the pressure feedback model examined here (Eq.(12))(Note that the unit of age of the universe is used in reference to the response of Boltzmann to Zermelo quoted previously, but it is purely illustrative and not very rigorous since $ \tau $ is likely to have been much shorter in the primordial universe).} 
\label{tab:1} 
\begin{center}     
\begin{tabular}{lll}
\hline\noalign{\smallskip}
Macrostate & Model & Recurrence time  \\
\noalign{\smallskip}\hline\noalign{\smallskip}
\includegraphics[width=2cm]{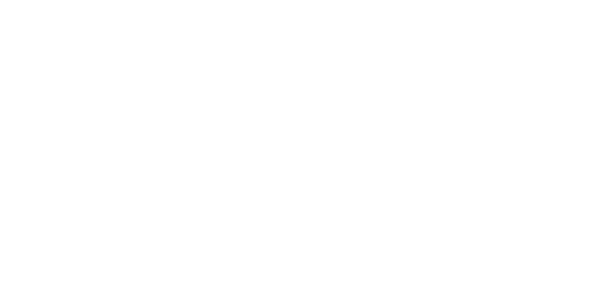}\includegraphics[width=2cm]{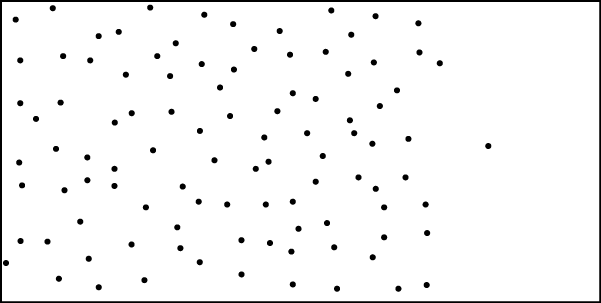} & Eq.(1) & 0.5 nanosecond  \\
$ \geq  $ 90\% particles in 3/4 of the box & Eq.(12) & 2 hours 45 minutes \\
\ & \ & \ \\
\includegraphics[width=2cm]{spacer.png}\includegraphics[width=2cm]{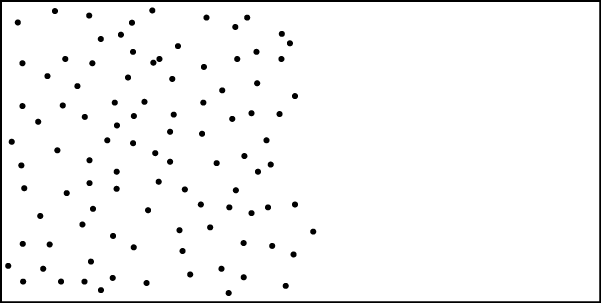} & Eq.(1) & 1 hour 47 minutes \\
$ \geq  $ 90\% particles in a half-box & Eq.(12) & $ 10^{47} $ universe ages \\
\noalign{\smallskip}\hline
\end{tabular}
\end{center} 
\end{table}
The system considered in Table 1 is very small since it corresponds to only 0.37 billionth of a picoliter of gas under one atmosphere, but it is however sufficient to reveal the astonishing gap between the predictions of Eq.(1) and Eq.(12). Even when using the time step used by Boltzmann ($ 0.5 \times 10^{-27} $ s) which is many times shorter than $ \tau  $ used here, the former calculation gives amazingly smaller recurrence times. This difference results from the introduction in the present study of a role, so far neglected, of pressure. When taking this role into account, recurrence times of Poincar\'e, even for very small systems, become purely mathematical but no longer physical, in agreement with the response of Boltzmann to Zermelo, thus restoring the idea of macroscopic irreversibility.

\section{Discussion}
The tool of discretization is used twice in this study: for space in the bulk approach, and for time for measuring recurrence times. The discretization of physical parameters previously conceived as continuous, long proved successful: (\textbf{i}) it allows to get rid of dimensions and to use unitless variables, (\textbf{ii}) it takes advantage of the powerful discrete probabilities and (\textbf{iii}) incidentally, it opened the way to quantum mechanics. As established for matter since Boltzmann, and for energy since Planck and Einstein, the continuous appearance of space and time could be also, after all, only illusions resulting from scale separation. Anyway, following the advice of Mark Kac, for finding a lot of solutions in mathematical physics, "Be wise, discretize!". The thermodynamic quantum of time $ h/4k_{B}T $ defined and used here is satisfactory enough in that it is universal, only dependent on temperature but not on parameters such as specific densities or collision rates. It could be very useful to make the connection between probabilities and time. It is important to precise that this time step is microscopic and reversible, valid as well in equilibrium and nonequilibrium situations. In this respect, it should not be confused with the thermodynamic arrow of time, which is macroscopic by nature and defined by the tendency of equilibration \cite{Lebowitz}. For example, the arrow of time, but not time steps, disappears for resting water in a bucket, in spite of its incessant microscopic movements. \\
The new formulations provided here show that the propensity of closed dynamic systems to homogenize is much stronger than previously anticipated, and highlight the reciprocal influences between individuals and the population, which are a fundament of statistical mechanics. This study formally quantifies and further supports the probabilistic view of Boltzmann without denying the existence of fluctuations. It also provides tools for explicitly quantifying the nonphysical status of the Poincar\'e recurrence theorem. According to this theorem, after sufficiently long time, any finite system can turn into a state which is very close to its initial state, which means that the second law of thermodynamics will be broken in even macroscopic scale. But only brains of mathematicians can be not impressed by the value of the "sufficiently long time" given by Eq.(12) in Table 1, which tremendously exceeds the lifetime of any closed system in the evolving universe. Mathematical modeling of physical phenomena, though incomplete, is certainly one of the noblest successes of humankind, but care should be taken, particularly in the field of probabilities, to always distinguish between mathematics and physics, following the recommendation of Boltzmann. For instance, certain asymptotes which are never reached from a mathematical viewpoint, are reached in the real world. Conversely, infinitesimal probabilities which exist mathematically, must be translated as "never" in the real world.

\end{document}